\documentclass[conference]{IEEEtran}
\IEEEoverridecommandlockouts
\usepackage{cite}
\usepackage{amsmath,amssymb,amsfonts}
\usepackage{algorithmic}
\usepackage{float}
\usepackage{booktabs}
\usepackage{titlesec}

\usepackage{graphicx}
\usepackage{textcomp}
\usepackage{xcolor}
\def\BibTeX{{\rm B\kern-.05em{\sc i\kern-.025em b}\kern-.08em
    T\kern-.1667em\lower.7ex\hbox{E}\kern-.125emX}}
\begin{document}

\title{Bi-LSTM Price Prediction based on Attention Mechanism\\}

\author{\IEEEauthorblockN{1\textsuperscript{st} Jiashu Lou}
\IEEEauthorblockA{\textit{College of Mathematics and Statistics} \\
\textit{Shenzhen University}\\
Shenzhen China \\
2019191138@email.szu.edu.cn}
\and
\IEEEauthorblockN{2\textsuperscript{nd} Leyi Cui}
\IEEEauthorblockA{\textit{College of Mathematics and Statistics} \\
\textit{Shenzhen University}\\
Shenzhen China \\
2019193024@email.szu.edu.cn
}
\and
\IEEEauthorblockN{3\textsuperscript{rd} Ye Li}
\IEEEauthorblockA{\textit{College of Economics} \\
\textit{Shenzhen University}\\
Shenzhen China\\
meloliye@126.com
}
}

\maketitle

\begin{abstract}
With the increasing enrichment and development of the financial derivatives market, the frequency of transactions is also faster and faster. Due to human limitations, algorithms and automatic trading have recently become the focus of discussion. In this paper, we propose a bidirectional LSTM neural network based on attention mechanism, which is based on two popular assets, gold and bitcoin. In terms of Feature Engineering, on the one hand, we add traditional technical factors, and at the same time, we combine time series models to develop factors. In the selection of model parameters, we finally chose two-layer deep learning network. According to AUC measurement, the accuracy of bitcoin and gold are 71.94\% and 73.03\% respectively. Using the forecast results, we achieved a return of 1089.34\% in two years. At the same time, we also compare the attention Bi-LSTM model proposed in this paper with the traditional model, and the results show that our model has the best performance in this data set. Finally, we discuss the significance of the model and the experimental results, as well as the possible improvement direction in the future.
\end{abstract}


\section{Introduction}
\subsection{Price prediction based on machine learning}
The traditional perception of the stock price is mainly based on technical analysis and fundamental analysis. However, with the development of big data, people increasingly realize that these two methods do not make full use of market public information. Therefore, asset price prediction based on machine learning has become a hot topic recently.\par
Adebiyi A. Ariyo, etc. predict the stock price based on the ARIMA model, which has been proven to have strong short-term prediction potential and can compete with the existing stock price prediction technology\cite{9}. Carson Kai-Sang Leung etc. proposed structural support vector machines (SSVMs) and use SSVM to predict the positive or negative changes of its share price\cite{10}. Recently, with the continuous development of deep learning technology and computing hardware, stock price prediction based on deep learning has become a hot topic. Sreelekshmy Selvin, etc. based on LSTM, CNN, and RNN, the stock price prediction experiment is carried out, and a sliding window is proposed to predict the short-term future value\cite{11}. Other works use stochastic process to predict a time series\cite{12}.

\subsection{Our work}
This article is based on two popular assets - gold and bitcoin, which are considered to be good hedging assets. We will only make predictions based on the price series itself, although this is not common, because stock analysts usually refer to a large number of factors. We will focus on the application of neural networks in time series, and hope to expand it to general time series data sets.\par

To provide an effective trading strategy, we prefer a prediction model and a programming model to tackle problems.\\
\indent Since the only data we can use to formulate the strategy is the daily prices of assets over five years. We first conduct feature engineering to excavate attributes through economical and mathematical methods. A sliding window is also introduced to lengthen the training set for the following prediction.\\
\indent Then we constantly improve the model based on the basic LSTM model, combining various optimize algorithms and mechanisms. After proving that our At-BiLSTM model is accurate, we calculate the effective duration of the model and predict the rate of return based on the duration.\\

\section{Data pre-processing}
To preserve the effectiveness of prediction and strategies, the continuity and authenticity of the trading data must be guaranteed. However, not all the data we have are integral. To ameliorate the condition of the data set, three methods have been proposed to improve the data as shown below:
\begin{itemize}
\item For transaction date, we convert different time forms into the same data type.
\item On days the market is closed, we take the gold price of the previous trading day within the available time as the price of that day, to obtain the corresponding prices of the two assets on each trading day within five years.
\item While training the model, we will set up a sliding window and continuously make predictions with the latest available historical data as input information.
\item Lagrange interpolation method is used for data padding.
\end{itemize}
\subsection{Lagrange interpolation method}
The Lagrange interpolation method is a polynomial interpolation method. The method selects several suitable values around the interpolation point to construct a simple interpolation function $y=G_j{(x)}$, which goes through the interpolation points. The simple but accurate interpolation function $G_j{(x)}$ utilizes the existing data and computes the corresponding data accordingly.
\begin{equation}
L(x)=\sum_{j=0}^l G_j(x)y_j
\end{equation}
where $G_j(x)$ is the weighting coefficient function.
\begin{equation}
G_j{(x)}=\prod_{k=0,k\neq j}^l \frac{x-x_k}{x_j-x_k}=\frac{x-x_0}{x_j-x_0}\cdots \frac{x-x_{j-1}}{x_j-x_{j-1}}
\frac{x-x_{j+1}}{x_j-x_{j+1}}
\end{equation}
We fill in the missing prices of gold over the five years using python.
\subsection{Feature engineering}
Since the only data we can use are the daily prices of two assets over five years, we conduct feature engineering to acquire more attributes to be used for analysis. By decomposing and aggregating raw data, we can better analyse the essence of the problem. In this part, two methods are applied to unearth new clusters of attributes.
\subsubsection{Economical attributes}
Referring to the real financial market, we calculate the following eight economical attributes:
\begin{itemize}
\item \textbf{Rate of return:} Measure the price change over two days by the equation below:
\begin{equation}\label{shou}
Return_t=\frac{P_t-P_{t-1}}{P_t}
\end{equation}
For the convenience of prediction, we will predict the daily rate of return in the following section \ref{Predict}.
\item \textbf{Price variance:} Measure the deviation of asset prices. We generate 10-day variance and 20-day variance.
\item \textbf{Moving average index:} Measure the average asset prices over a period of time. We generate 10-day MA and 30-day MA.
\item \textbf{Boll index:} Created by Mr.John Boll,\cite{2}
 who uses statistical principles to find out the standard deviation and confidence interval to determine the volatility range and future trend of the stock price. We generate high, middle and low boll value through the equation below:
\begin{equation}
Boll_t = \frac{\sum\limits_{n=1}^{N}p_{t-n}}{N}\pm 2\sigma_B
\end{equation}
where $\sigma_B$ is the price standard deviation.
\item \textbf{Psychological index:} Measure the sentiment index of investors' psychological fluctuations in the asset market.
\begin{equation}
Psy_t=\frac{N_{up}}{N}
\end{equation}
where $N_{up}$ is the number of days during the last $N$ days.
\item \textbf{Technical indicator:} We also introduce a popular indicators, RSI, in financial markets to expand our economical attributes. With reference to the definitions of the indicator in the stock market, we can calculate the corresponding asset attribute.
\begin{equation}
RSI=100-\frac{100}{1+(\frac{1}{n}\sum\limits ^n_{i=0}r_{t-i})/(\frac{1}{n}\sum\limits ^n_{i=0}f_{t-i})}
\end{equation}
where $r_{t-i}$ is the total change in price rising days and $f_{t-i}$ is the total change in price falling days.
\end{itemize}
\subsubsection{Mathematical attributes}
Empirical research shows that most financial data have the characteristics of sharp peaks and thick tails, so they cannot be characterized by normal distribution. The independence assumption of the sequence is generally no longer valid and the prices of financial assets present autocorrelation and volatility cluster in time series. To capture the asymmetrical data, we choose GARCH model to describe the conditional heteroskedasticity over time series and introduce attributes:
\begin{itemize}
\item \textbf{Stochastic disturbance term $\mu_t$}
\item \textbf{Conditional variance $\sigma^2_t$}
\end{itemize}

\begin{figure*}[H]
\centering
\begin{minipage}[t]{0.5\linewidth}
\centering
\includegraphics[width=1\linewidth]{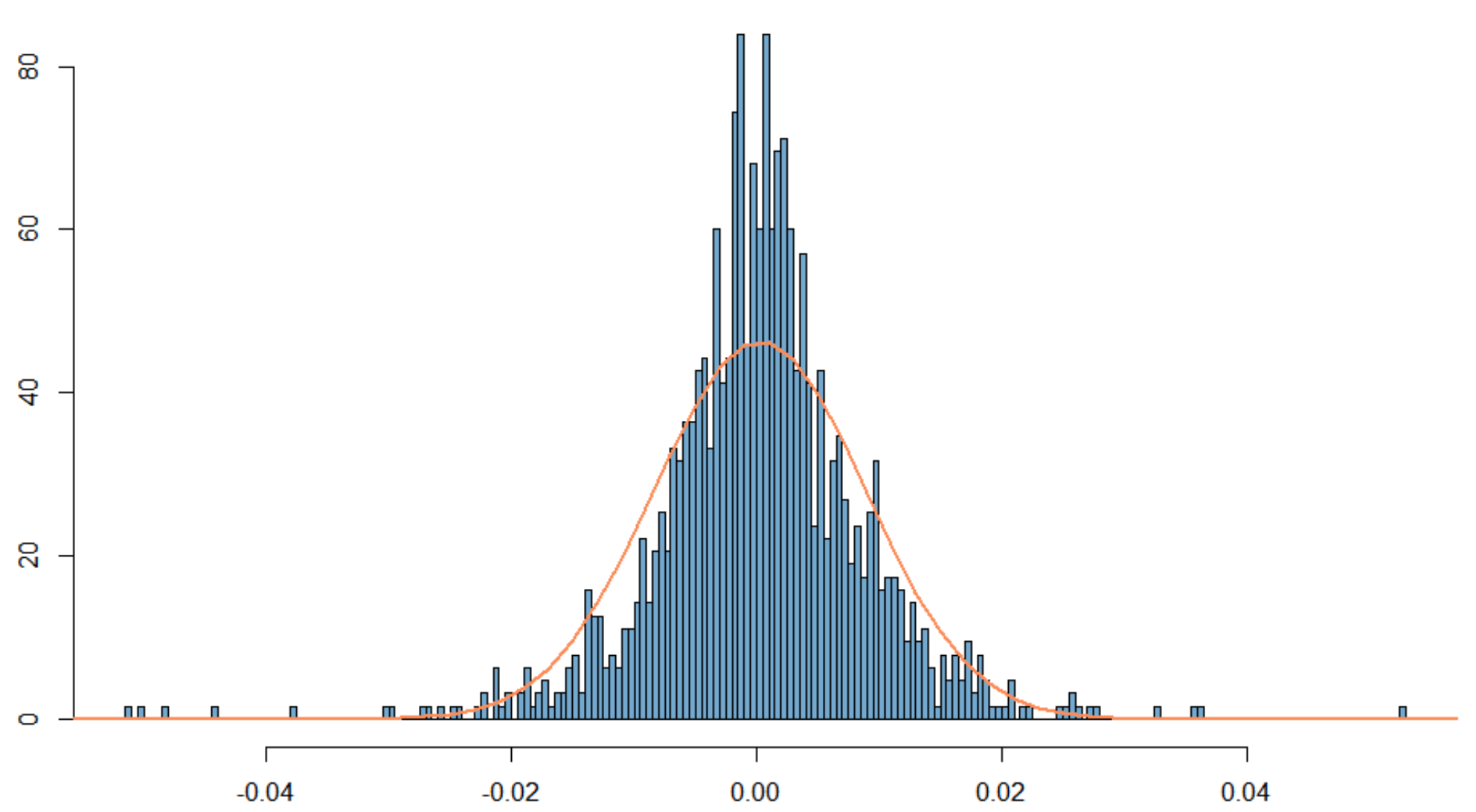}
\caption{Price return of gold}
\end{minipage}
\begin{minipage}[t]{0.5\linewidth}
\centering
\includegraphics[width=1\linewidth]{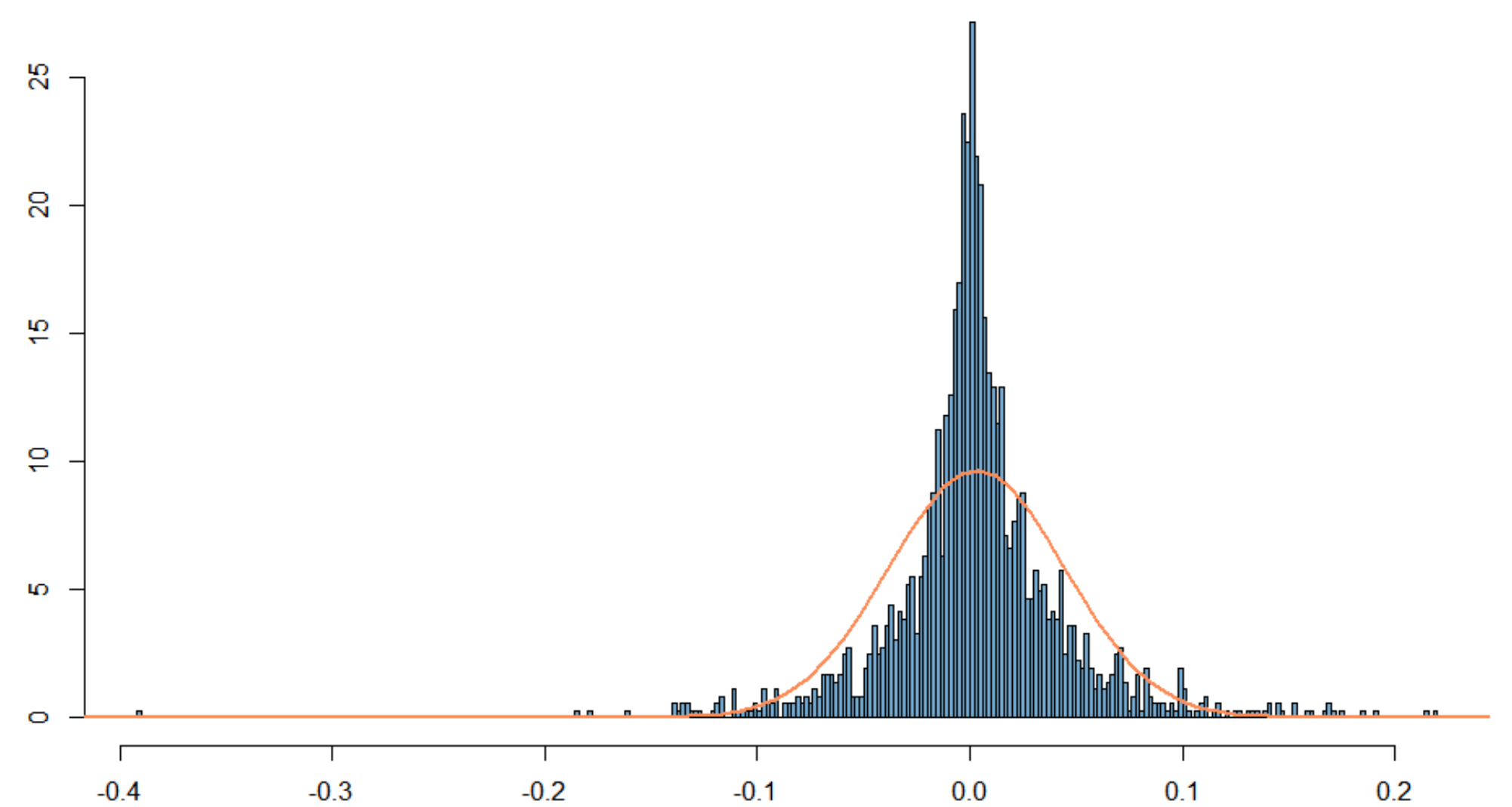}
\caption{Price return of bitcoin}
\end{minipage}
\end{figure*}

GARCH model is a widely used time series analysis method proposed by Bollerslev (1986), which are an extension of the earlier work on ARCH models by Engle (1982).\cite{1}
The core idea of ARCH model is that the variance of error at time $t$ depends on the previous variance of error.
\begin{equation}
\left\{
\begin{array}{l}
y_t=X^{`}_t\phi+\mu_t, \mu_t \sim N(0,\sigma^2_t)\\
\sigma^2_t=\alpha_0+\sum\limits_{i=1}^p\alpha_i\mu^2_{t-i}
\end{array}
\right.
\end{equation}
where $\mu_t$ represents the stochastic disturbance term without serial correlation and $\sigma^2_t$ represents the variance of the stochastic disturbance.\\
\indent Then we generate GARCH model by adding the lag of $\sigma_t$ to ARCH model:
\begin{equation}
\left\{
\begin{array}{l}
y_t=X^{`}_t\phi+\mu_t, \mu_t \sim N(0,\sigma^2_t)\\
\sigma^2_t=\alpha_0+\sum\limits_{i=1}^p\alpha_i\mu^2_{t-i}+\sum\limits_{i=1}^q\beta_i\sigma^2_{t-i}
\end{array}
\right.
\end{equation}
where $q$ and $p$ represent the lags in the GARCH term and the ARCH term respectively.\\
\indent In order to use the GARCH model, we need to verify the stationary time process and normality of the acquired data before extracting features. Due to the huge amount of data, the ADF method is used in the test.
\begin{table*}[htbp]
\centering
\caption{ADF test}
\begin{tabular}{c|cccccccc}
\hline
\textbf{Asset}        & \multicolumn{4}{c}{\textbf{gold}}                             & \multicolumn{4}{c}{\textbf{bitcoin}}                          \\ \hline
\multicolumn{1}{l|}{} & \textbf{tau1} & \textbf{tau3} & \textbf{phi2} & \textbf{phi3} & \textbf{tau1} & \textbf{tau3} & \textbf{phi2} & \textbf{phi3} \\ \hline
\multicolumn{1}{l|}{} & -29.3978      & -29.665       & 293.3379      & 440.0067      & -23.5781      & -23.6103      & 185.8161      & 278.724       \\
\textbf{1pct}         & -2.58         & -3.96         & 6.09          & 8.27          & -2.58         & -3.96         & 6.09          & 8.27          \\
\textbf{5pct}         & -1.95         & -3.41         & 4.68          & 6.25          & -1.95         & -3.41         & 4.68          & 6.25          \\
\textbf{10pct}        & -1.62         & -3.12         & 6.25          & 5.34          & -1.62         & -3.12         & 6.25          & 5.34          \\ \hline
\end{tabular}
\end{table*}
ADF test shows that at the $1\%$ confidence level, we can reject the null hypothesis and consider the series of gold and bicoin as stable.\\
\indent GARCH(1,1) model has few parameters and is often tested in practice. Since it is quite difficult to determine the order of the GARCH model, so we choose commonly used GARCH(1,1) model to extract attributes. From GARCH(1,1), we acquire two attributes $\mu_t$ and $\sigma_t^2$, referring  to stochastic disturbance team and conditional variance respectively.
\begin{equation}
\mu_t=y_t-x_t\phi
\end{equation}
\begin{equation}
\sigma^2_t=\alpha_0+\alpha_1\mu^2_{t-1}+\beta_1\sigma^2_{t-1}
\end{equation}
where $y_t$ is the asset price in time $t$, $x_t$ is the asset price in time $t-1$, $\mu_t$ is the stochastic disturbance term, $\sigma^2_t$ is the variance.\\
\begin{figure*}[htbp]
\centering
\begin{minipage}[t]{0.5\textwidth}
\centering
\includegraphics[width=1\textwidth]{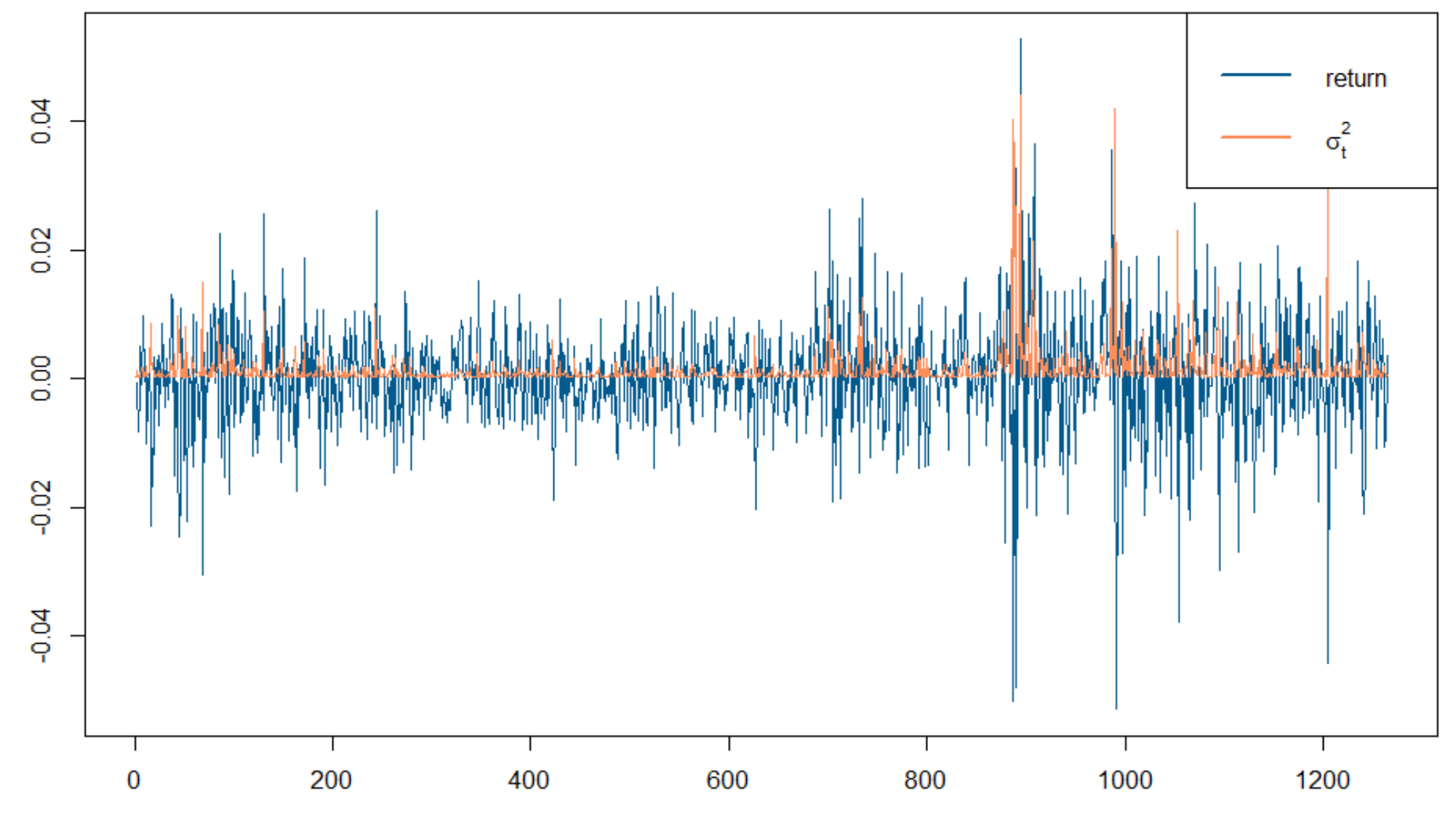}
\end{minipage}
\begin{minipage}[t]{0.5\textwidth}
\centering
\includegraphics[width=1\textwidth]{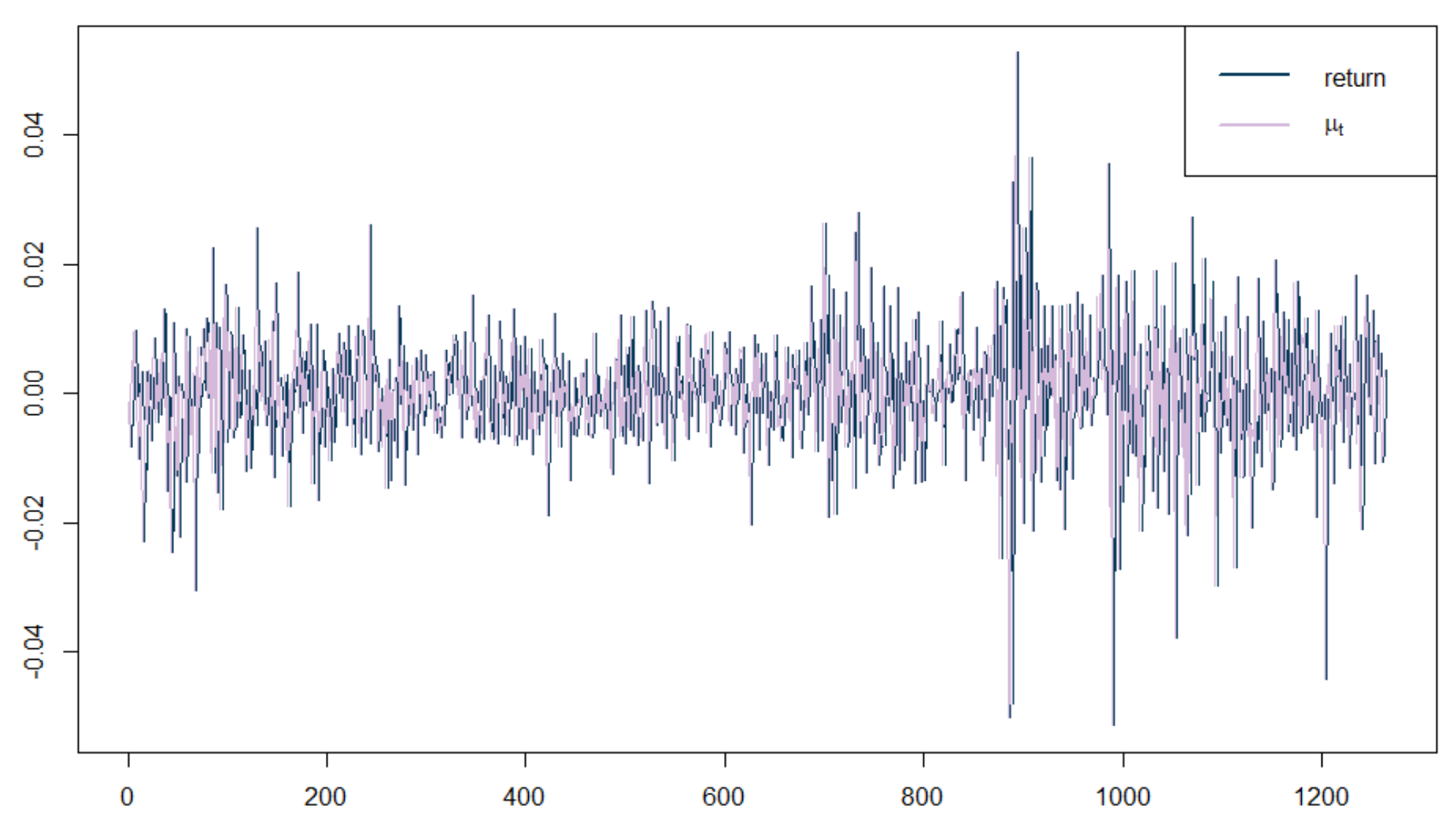}
\end{minipage}
\caption{Attribute distribution of gold}
\end{figure*}

\begin{figure*}[htbp]
\centering
\begin{minipage}[t]{0.5\linewidth}
\centering
\includegraphics[width=1\textwidth]{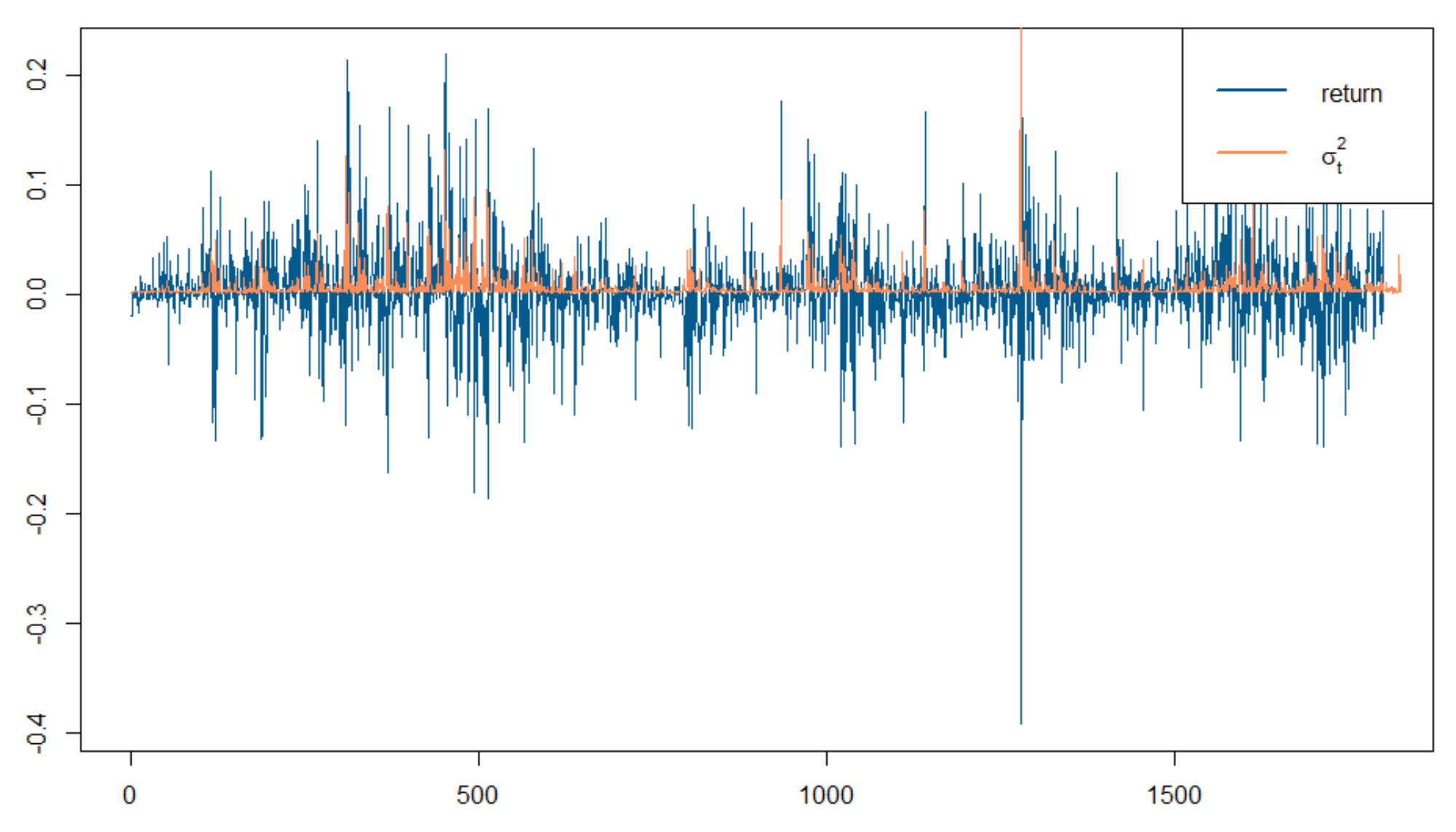}
\end{minipage}
\begin{minipage}[t]{0.5\linewidth}
\centering
\includegraphics[width=1\textwidth]{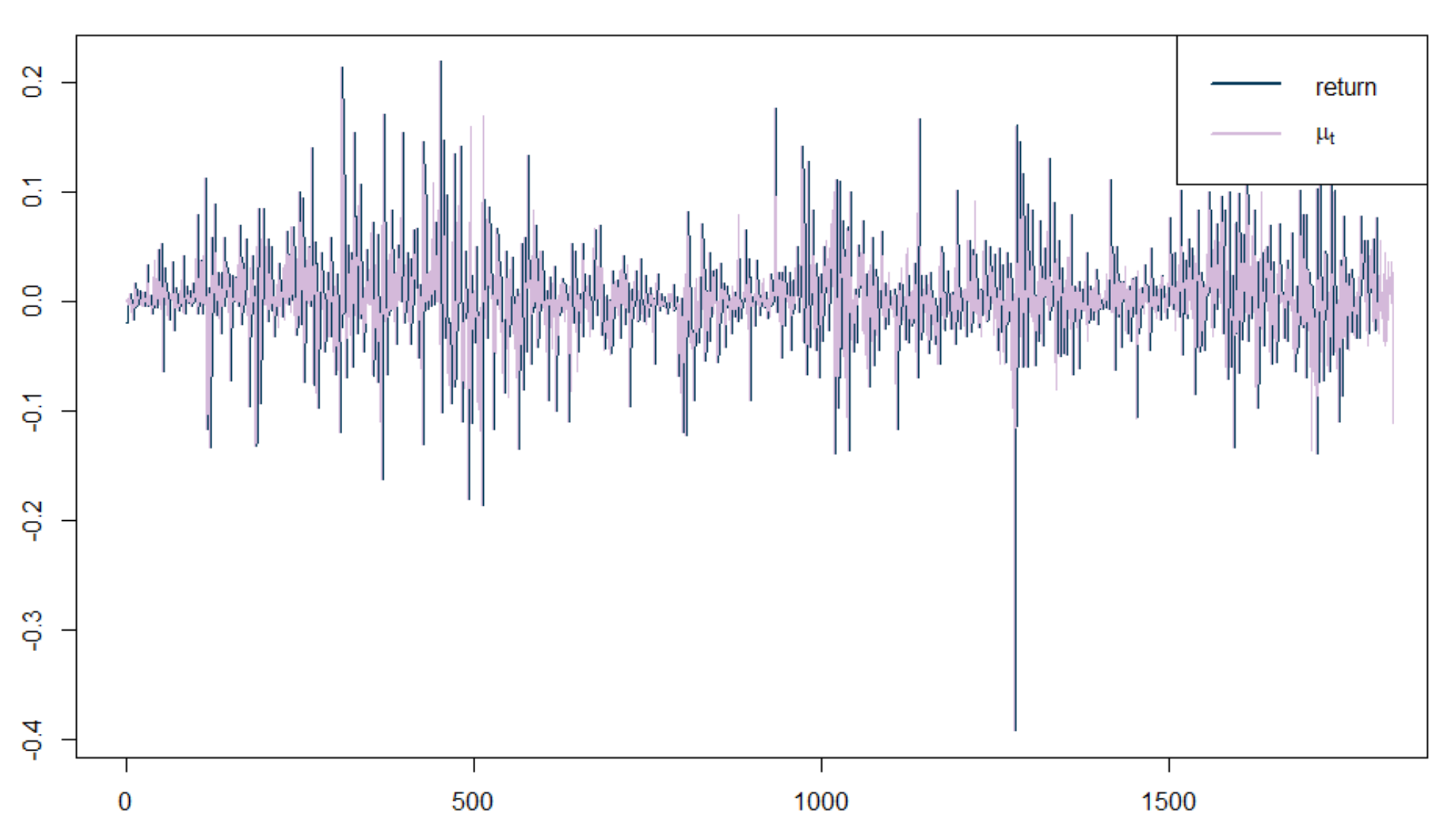}
\end{minipage}
\caption{Attribute distribution of bitcoin}
\end{figure*}
As the figure shown, attributes $\mu_t$ have a high degree of coincidence with asset return and attributes $\sigma^2_t$ are all positive. So they are suitable for describing asset return.
\section{Prediction model}\label{Predict}
Since there are many factors affecting the volatility of stock prices, deep learning models such as neural networks can be used for prediction. The most popular model used for time series prediction is RNN model. The chain-type structural feature of RNN is particularly suitable to process the discrete data series. Since the information of hidden layer comes only from the information of current input and last layer, however, there is no memory function in RNN model. It leads to gradient disappearance problems. In this context, LSTM (Long-short term memory) model has emerged.
\subsection{LSTM model}
Cells were introduced to LSTM model, which utilizes three gates, forget, input and output, to maintain and control information. The forget gate combines the information of the previous hidden layer $s_{t-1}$ and input $x_{t}$. Under the sigmoid function, it decides whether the old information to be discard or not. The input gate and tanh function filter the activation information and work out new cell $\tilde{c_{t}}$. The output gate finally output the information after processing. In the model, the decay propagation of the gradient information can be solved to keep the network in memory for a relative long time.
	\begin{figure}[htbp]
	\centering
	\includegraphics[width=\linewidth]{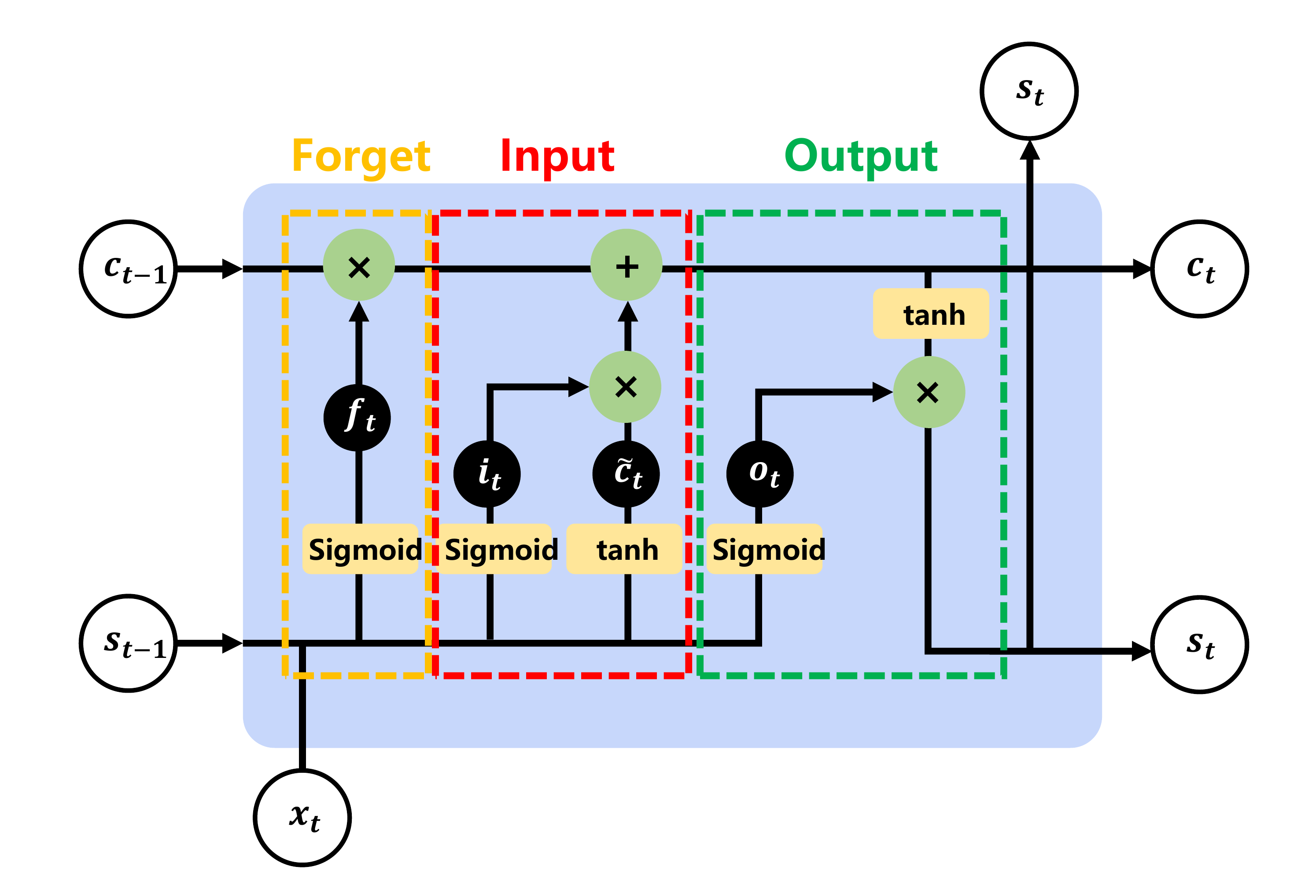}
	\caption{The structure of LSTM}
	\end{figure}
The structure of the "gate" realizes the addition and removal of information, and determines whether the model will save or forget the information during training. With this "gate structure", LSTM model effectively alleviates the problem of gradient change and selectively stores information. For a given feature at $t$, the network will have the ability to backward predict the target information, enabling investors to predict the future price at some time, so as to make more accurate decisions.
\subsubsection{Two-layer LSTM with Batch Normalization}
The research conducted by Li(2021)\cite{7} shows that the structure of two-layer LSTM can effectively improve the accuracy of the model prediction and the study presented by Sergey Ioffe(2015)\cite{5} shows that Batch Normalization could accelerate deep Network Training by Reducing Internal Covariate Shift. So we make some improvement in the traditional model correspondingly.\\
\indent Batch Normalization is an efficient data normalization method used to activate layers in deep neural networks. It can speed up the convergence rate of the model training, make the training process more stable, avoid gradient explosion or gradient disappearance and play a certain regularization role.
\begin{equation}
\begin{array}{l}
Input:B=\{x_{1\cdots m}\}\quad \gamma,\beta(parameters)\\
Output:\{y_i=BN_{\gamma,\beta}(x_i)\}\\
\mu_B \gets \frac{1}{m}\sum\limits^m_{i=1}x_i\\
\sigma^2_B\gets \frac{1}{m}\sum\limits^m_{i=1}(x_i-\mu_B)^2\\
\tilde{x_{i}}\gets \frac{(x_i-\mu_B)^2}{\sqrt{\sigma^2_B+\varepsilon}}\\
y_i\gets \gamma \tilde{x_{i}}+\beta
\end{array}
\end{equation}
where B is the set of input values, a learnable parameter, which can be used to normalize with and restore the data.\\
\indent The data is normalized to a unified interval to reduce the data divergence and the learning complexity of the network. After normalization, using $\gamma$ and $\beta$ as the parameters preserves the distribution of the original data. To further avoid overfitting, we add "Dropout layers" in the model. The layers can temporarily discard some neural from the network during the training process of a deep learning network refereed to the probability. The mechanism helps find sparse network containing only a fraction of the neurons  based on the original complex network, greatly reducing the possibility of overfitting.\cite{6}

\subsection{Bidirectional LSTM model}
To further improve the efficiency of data use and enhance the robustness of the model, we combine a forward LSTM model with a backward LSTM model to construct Bidirectional LSTM model. Bi-LSTM can effectively use the forward and backward feature information of the input and improve the accuracy when the parameters such as the training rounds are unchanged.
\subsection{Attention mechanism}
When assessing a set of information, our neurons automatically scan the global image and then focus on the target based on past experience and current goals, that is, the focus of attention. Once the focus is determined, we pay more attention to more details in that area and suppress other useless information. For example, in the field of image processing, when scanning journal images, people mainly focus on the picture and the title, in line with our life experience.\\
\indent Weight factors are assigned to each factor to measure its contribution in attention mechanism. Firstly, two hidden layer, Encoder and Decoder, equivalent to the input and output of the structure are employed. Points are then accumulated with the Decoder hidden layer and each Encoder hidden layer. The results are recorded as "Score". All scores are then sent to the softmax layer so that the higher score the layer achieves, the greater the probability would be. Thus the mechanism suppresses invalid or noise information.
\begin{figure}[htbp]
\centering

\centering
\includegraphics[width=0.6\linewidth]{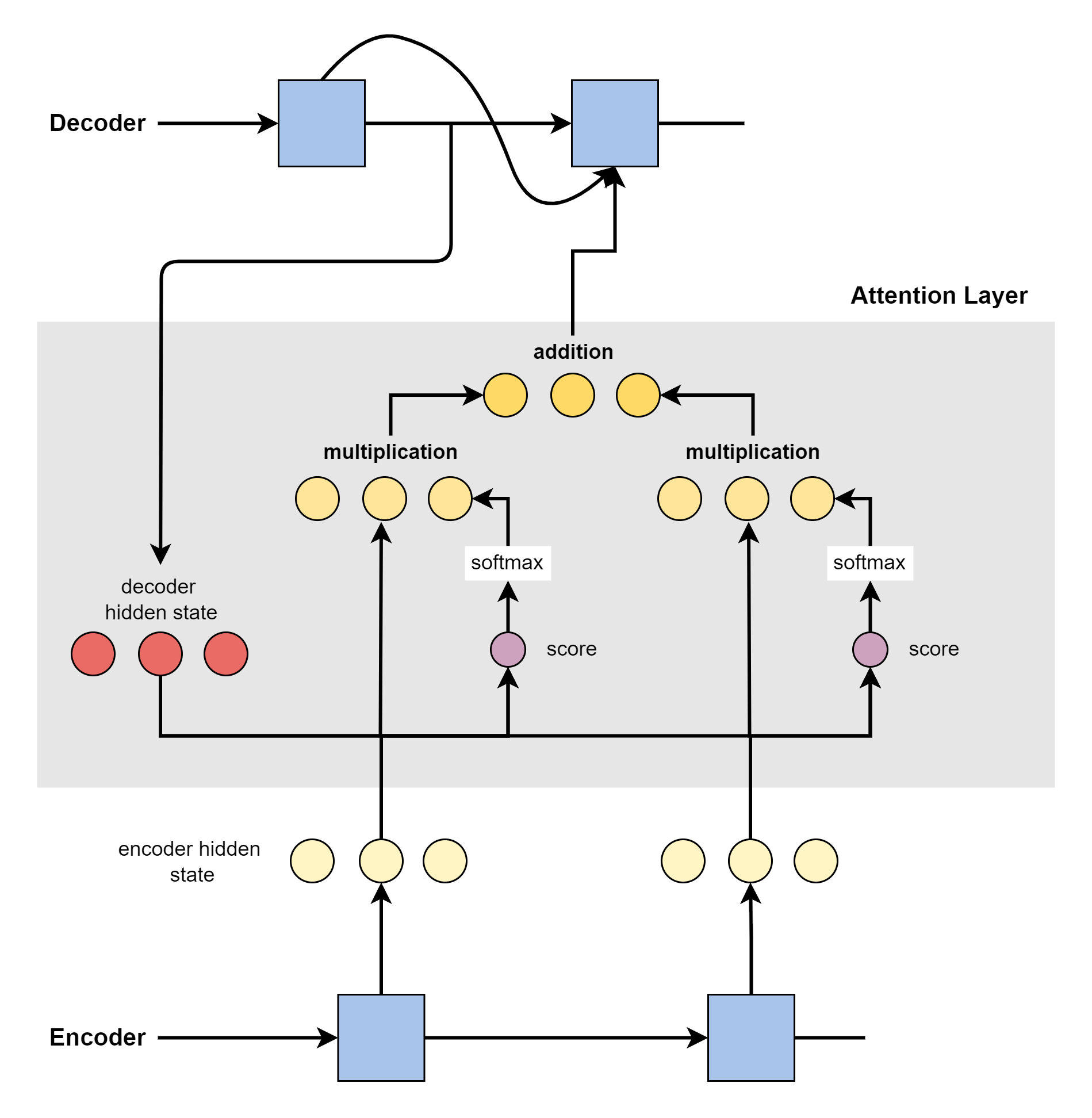}
\caption{Structure of attention mechanism}
\end{figure}
\begin{figure}

\centering
\includegraphics[width=0.6\linewidth]{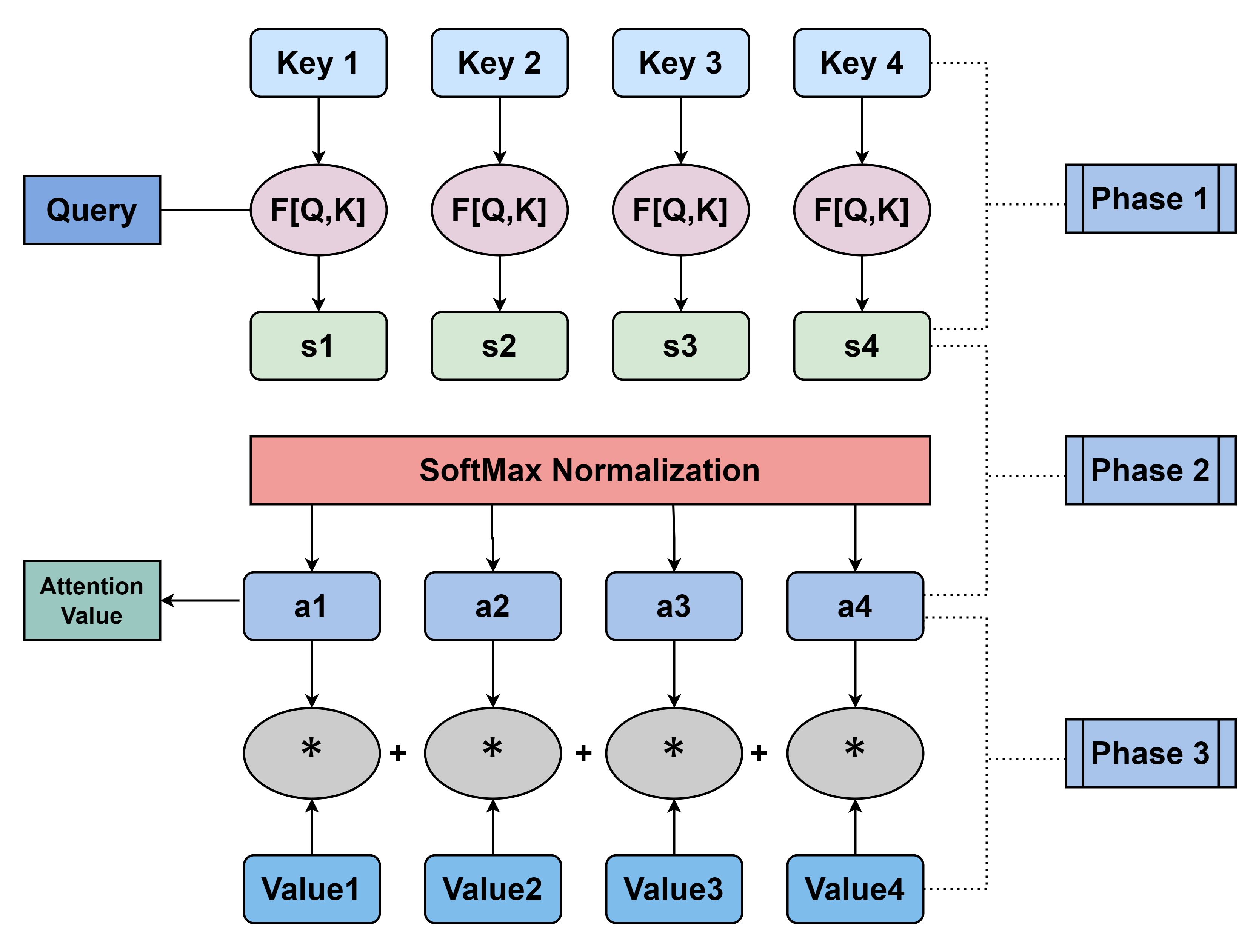}
\caption{Process of attention mechanism}

\end{figure}
where the softmax function is:
\begin{equation}
softmax(z)_i = \frac{exp(z_i)}{ {\textstyle \sum_{j}^{}}exp(z_j) }
\end{equation}
\indent For time series data, due to high-dimensional variables and multi-step time steps, if trained with equal weights, it will lead to unclear information focus, eventually resulting in overfitting noise information. Therefore, we introduce attention mechanism to the our model. We apply attention mechanism in the time step dimension of the input variable, considering the past trading data as significant factors during the prediction of future return.
\subsection{Results}
\subsubsection{RMSProp optimization algorithm}
While training the model, we choose RMSProp optimization algorithm to adjust model parameters.
\begin{equation}
g\gets \frac{1}{m} \bigtriangledown_\theta \sum\nolimits_i L(f(x^{(i)};\theta),y^{(i)})
\end{equation}
\begin{equation}
r_{t}\leftarrow \rho r_{t-1}+(1-\rho)g\odot g
\end{equation}
\begin{equation}
\Delta\theta=-\frac{\epsilon}{\sqrt{\delta+r}}\odot g
\end{equation}
where $\rho$ is used to control how much historical information is acquired.\\
\indent Given that the loss functions of neural networks are all non-convex, RMSProp\cite{3} presents better in non-convex conditions, changing the cumulative gradient to the exponential decay moving average to discard distant past history.
\indent Based on the above analysis, we finally determined the optimal parameter portfolio as follows:
\begin{table}[htbp]
\centering
\caption{Optimal parameter portfolio}
\begin{tabular}{@{}c|cccc@{}}
\toprule
                   & \textbf{Units} & \textbf{batch\_size} & \textbf{Learning rate} & \textbf{epochs} \\ \midrule
\textbf{LSTM}      & 128             & 128                  & 0.01                   & 300             \\
\textbf{BiLSTM}    & 64             & 128                  & 0.001                  & 300             \\
\textbf{At-BiLSTM} & 32            & 128                  & 0.01                   & 300             \\ \bottomrule
\end{tabular}
\end{table}
We input two assets datasets into the three models respectively and compare the accuracy of the three models between the training set and the test set.
\begin{figure}[htbp]
\centering
\includegraphics[width=0.5\textwidth]{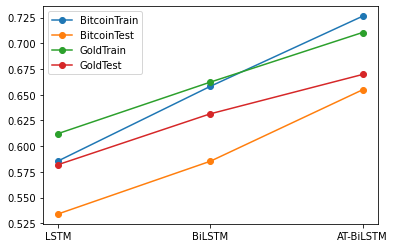}
\caption{Accuracy of three models}
\end{figure}
The accuracy of the three models gradually increase for both assets, indicating that AT-BiLSTM model is prominent in prediction. In addition to the above three models, we compared the accuracy of several common prediction models and obtained the results as follows:

Considering the above analysis, we finally decide to use the AT-BiLSTM model with high predictive ability for return prediction during the asset trading period. Using the equation \ref{shou}, we can work out a five-year forecast for the prices of two assets. In order to show the difference between prediction and reality more directly, we also work out the local price trend.
\begin{figure*}[htbp]
\centering
\begin{minipage}[t]{0.5\linewidth}
\centering
\includegraphics[width=1\linewidth]{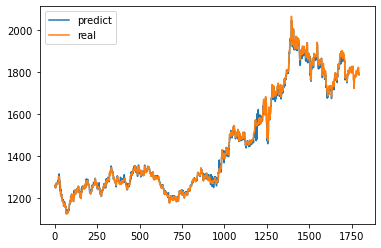}
\caption{Prediction of Gold}
\end{minipage}
\begin{minipage}[t]{0.5\linewidth}
\centering
\includegraphics[width=1\linewidth]{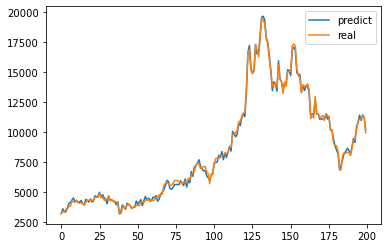}
\caption{Prediction of Bitcoin}
\end{minipage}
\end{figure*}
\section{Firm offer training strategy}
The prediction of return by the above models is a global train, assuming that we have all the trading data during the period. However, when we develop strategies, we cannot know the "future" data. For example, we suppose the trader were at 9/11/2016, the beginning of the dataset. At this point, the trader would not have any trading data, namely that there is no data for the trader to train the model. That means we must take out a considerable trading period at the beginning, during which we temporarily do not participate in the transaction. Only if the accuracy of the model rises to a certain extent, can we use the model to predict and make investment decisions. Therefore, we should divide the five-year period into training and trading periods.\\
\indent To save the training time, we update the model using an incremental training pattern. Taking gold trading as an example, the training error decreases over time. As the figure shown above, after 300 days the error plateaus, indicating the model is fully learned. So we select the $300^{th}$ day of the trading period as the point we start our strategy. When we start our strategy, we still train the prediction model with incremental training pattern, so that the model always learns the latest data. As can be seen from the attention mechanism above, the latest trading data presents greater learning value.\par
We also compared the attention bilstm model with other traditional models. Finally, the experimental and prediction results are shown in the TABLE III. It can be seen that the attention bilstm model achieves the best prediction effect. It also can be seen in the Fig 9 and Fig 10, that the model has a good fit for the price.

\begin{table*}[htbp]
\caption{accuracy of common models}
\centering
\begin{tabular}{@{}cccccc@{}}
\toprule
\textbf{Indicator}                 & \textbf{Assets}  & \textbf{LinearModel} & \textbf{SVM}     & \textbf{Decision tree} & \textbf{Arima Model} \\ \midrule
{\textbf{Acc.(\%)}} & \textbf{Bitcoin} & 0.5646               & 0.5505           & 0.5489                 & 0.5316               \\
                                   & \textbf{Gold}    & 0.5723               & 0.5589           & 0.5501                 & 0.5469               \\ \midrule
{\textbf{AUC}}      & \textbf{Bitcoin} & 0.6241               & 0.5004           & 0.5208                 & \textless{}0.5       \\
                                   & \textbf{Gold}    & 0.6405               & \textless{}0.5   & \textless{}0.5         & \textless{}0.5       \\ \midrule
\textbf{Indicator}                 & \textbf{Assets}  & \textbf{LSTM}        & \textbf{AT-LSTM} & \textbf{BiLSTM}        & \textbf{AT-BiLSTM}   \\ \midrule
{\textbf{Acc.(\%)}} & \textbf{Bitcoin} & 0.5357               & 0.6239           & 0.5854                 & 0.6578               \\
                                   & \textbf{Gold}    & 0.5821               & 0.6433           & 0.6315                 & 0.6697               \\ \midrule
{\textbf{AUC}}      & \textbf{Bitcoin} & 0.6755               & 0.7016           & 0.6988                 & 0.7194               \\
                                   & \textbf{Gold}    & 0.6817               & 0.7132           & 0.7102                 & 0.7303               \\ \bottomrule
\end{tabular}
\end{table*}

\section{Conclusion, Discussion and Future Work}
This paper have developed an automatic model that can predict the price, judge the trading point and determine the trading volume. This model is based on Bi-LSTM and adds attention mechanism. The timing and volume strategy based on dynamic programming and entropy weight method is also added to make trading decision. After our test, it has good prediction accuracy and strong robustness. In our simulation back test, it has achieved an annualized yield of $170\%$. The value of $\$1000$ investment would be $\$8542.3$. \par
Although our strategy can achieve high annual yield, ironically, its yield is not even as good as buying bitcoin at the beginning of the trading period and holding it all the time. This also shows that the model performs generally for one-sided markets. Besides, Although the attention Bi-LSTM model has been proved to be the best in our experiments, it can be seen that the performance gap between it and other models is not large. There is a possibility that the attention Bi-LSTM model does not perform well in other data sets, which also proves that there is no free lunch theorem\cite{8} from another side.\par
This paper only considers the prediction of price, and does not explore in depth how to place an asset order. For example, when our model predicts that prices will rise tomorrow, how many assets should we buy? At the same time, we also did not consider the matching of trading books, liquidity and other issues that may occur in the real trading process. These problems need to be further studied.

\end{document}